\documentstyle[12pt,psfig]{article}
\def\title#1{\relax\vspace*{2cm}{\large{\bf #1}}\par\vspace*{13.5pt}}
\def\author#1{{#1}\par\vspace*{13.5pt}}
\def\affil#1{{\it #1}\par}
\def\abstract{\vspace*{27pt}ABSTRACT\par\relax}
\def\section#1{\par{#1}\par}
\def\subsection#1{\par\underline{#1}\par}
\def\subsubsection#1{\par\underline{#1.}\ \ }
\def\acknow{\par ACKNOWLEDGMENTS\par}
\newenvironment{references}{\section{REFERENCES}\vspace*{.5cm}%
 \parindent=0pt\frenchspacing%
 \parskip=1pt plus 1pt minus 1pt%
 \interlinepenalty=1000\tolerance=400%
 \pretolerance=10000\hyphenpenalty=10000%
 \everypar={\hangindent=1.6pc}
}{}

\begin{document}

\title{AGN WATCH CONTINUUM MONITORING OF RADIO-QUIET AND
RADIO-LOUD AGN}
\author{Paul T. O'Brien$^{1}$, Karen M. Leighly$^{2}$}
\affil{$^{1}$Physics \& Astronomy Department, Leicester University,
University Road, Leicester, LE1 7RH, U.K.\\
$^{2}$Cosmic Radiation Laboratory, RIKEN, Hirosawa 2-1, Wako-shi, Saitama
351, Japan\\
$^{3}$Columbia Astrophysics Laboratory, 538 West 120th Street, New York,
NY 10027, USA
}

\abstract

The International AGN Watch has monitored a number of radio-quiet and
radio-loud Active Galactic Nuclei -- the most luminous objects in the
universe. We present a review of the main observational results from the
continuum monitoring campaigns, concentrating on those which aim to
quantify the simultaneous ultraviolet to X-ray variability
characteristics. These data provide strong constraints on the proposed
continuum emission mechanisms. The AGN Watch campaigns have made
extensive use of a wide variety of both ground- and space-based
multi-waveband observational facilities, and we stress that long-term
simultaneous access to the entire electromagnetic spectrum is essential
if further progress is to be made.

\section{THE INTERNATIONAL AGN WATCH}

In the late 1980's an informal consortium of about 100 astronomers was
formed to obtain large, high-quality multi-waveband data for
investigation of the continuum and emission-line variability
characteristics of Active Galactic Nuclei (AGN). The consortium was
formed in response to the realization that small groups were unlikely to
obtain large amounts of observing time, and would face great practical
difficulty in both coordinating observations over many months and rapidly
reducing data from numerous ground-based and orbiting observatories. The
AGN Watch exists to obtain such data and rapidly place them in the public
domain. Smaller collaborations, drawn from both inside and outside AGN
Watch, can then use these data to place firm constraints on the seemingly
ever-growing list of proposed AGN models. Here we summarize the continuum
monitoring campaigns. The emission-line monitoring is presented elsewhere
({\it e.g.} Peterson, these proceedings).

\begin{table}[ht]
{Table 1. \ \ AGN Watch Continuum Monitoring Campaigns\vfill}
\begin{tabular}{lll}
\noalign{\bigskip}
\hline
\hline
\noalign{\smallskip}
Object & Waveband (facility) & Data Publications\\
\noalign{\smallskip}
\hline
\noalign{\smallskip}
NGC5548 & UV \& Optical (IUE) & Clavel {\it et al.} (1991) \ (Paper I)\\
& Optical (Ground-based) & Peterson {\it et al.} (1991) \ (Paper II)\\
& & Peterson {\it et al.} (1992) \ (Paper III)\\
& & Dietrich {\it et al.} (1993) \ (Paper IV)\\
& & Peterson {\it et al.} (1994) \ (Paper VII)\\
& UV (HST) \& Optical (Ground-based) & Korista {\it et al.} (1995) \ (Paper VIII) \\
& EUV (EUVE) & Marshall {\it et al.} (1997)\\
\noalign{\smallskip}
NGC3783 & UV \& Optical (IUE) & Reichert {\it et al.} (1994) \ (Paper V)\\
& Optical (Ground-based) & Stirpe {\it et al.} (1994) \ (Paper VI)\\
\noalign{\smallskip}
NGC4151 & UV (IUE) & Crenshaw {\it et al.} (1996)\\
& Optical (Ground-based) & Kaspi {\it et al.} (1996)\\
& X-ray (ROSAT/ASCA) \& $\gamma$-Ray (CGRO) & Warwick {\it et al.} (1996)\\
& & Edelson {\it et al.} (1996)\\ 
\noalign{\smallskip}
Fairall 9 & UV (IUE) & Rodr\'{\i}guez-Pascual {\it et al.} (1997) (Paper
IX)\\
& Optical (Ground-based) & Santos-Lle\'o {\it et al.} (in prep.)\\
\noalign{\smallskip}
3C390.3 & UV (IUE) & O'Brien {\it et al.} (in prep.)\\
& Optical (ground-based) & Dietrich {\it et al.} (in prep.)\\
& X-ray (ROSAT/ASCA) & Leighly {\it et al.} (1997)\\
& Radio (MERLIN) & Leighly {\it et al.} (in prep.)\\
\noalign{\smallskip}
NGC7469 & UV (IUE/HST) \& X-ray (XTE) & Campaign June--July 1996\\
\noalign{\smallskip}
Mrk279 & IR (ISO) \& Optical (Ground-based) & Campaign 1996--1997 \\
\noalign{\smallskip}
\hline
         \end{tabular}
 \end{table}

\section{CONTINUUM MONITORING OF SEYFERT~1 GALAXIES}

The AGN watch has conducted campaigns on 7 AGN to date. The object names,
wavebands covered and publications in which the original data can be
found are given in Table~1. The initial campaigns concentrated on UV
(1150--3200\AA) and optical monitoring using IUE and numerous
ground-based telescopes. More recently the campaigns have been expanded
to the EUV, X-ray and $\gamma$-ray wavebands to investigate the
photoionizing continuum in detail, and to the IR to investigate the
properties of hot dust and the putative dense molecular torus around the
nucleus.

\begin{figure}[ht]
\psfig{figure=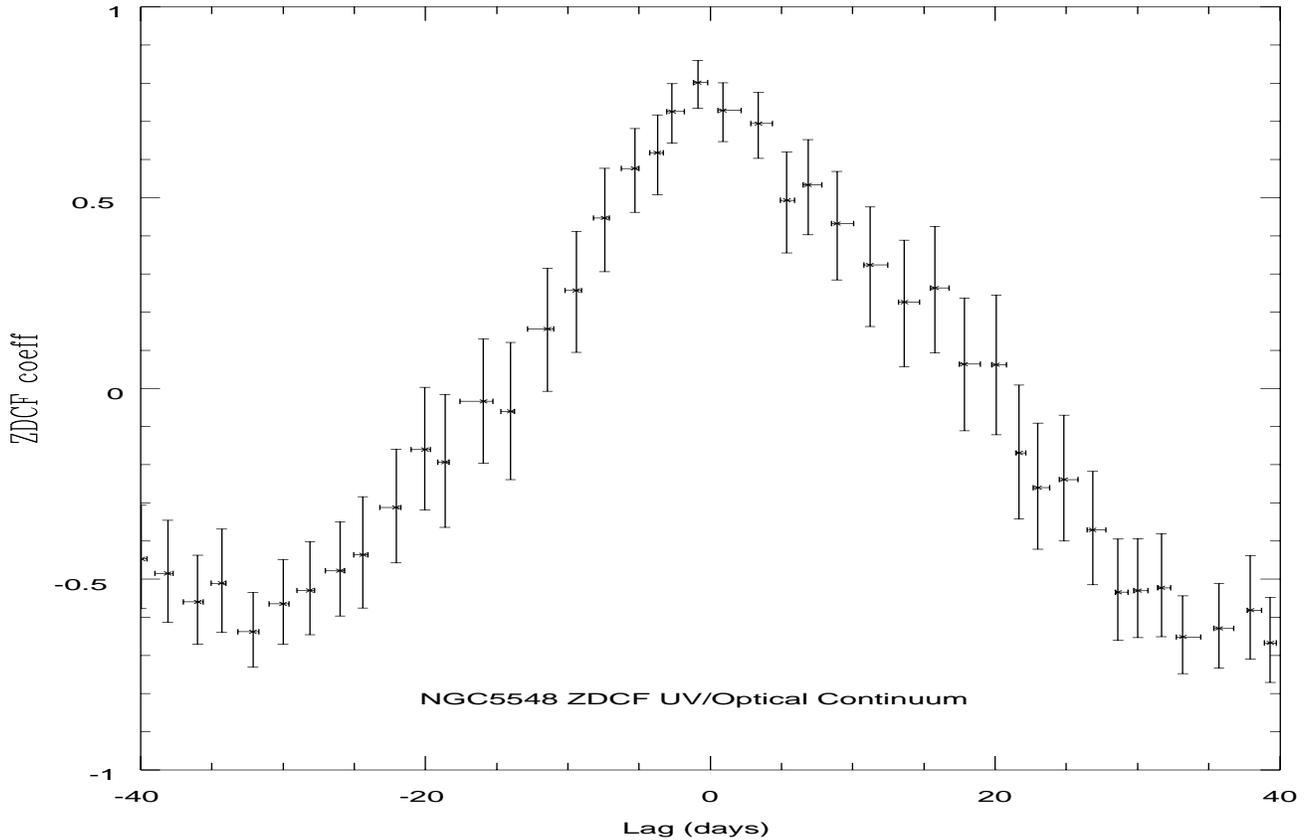,width=18cm,height=12cm,angle=0}
\caption{NGC5548 ZDCF computed between 1350 and 5100\AA\ using data
from Korista {\it et al.} (1995).\hfill}
\end{figure}

\subsection{UV and Optical Continuum Variability}

The individual UV and optical continuum variability characteristics of
the AGN monitored by AGN Watch are discussed in detail in the
publications listed in Table~1. The principal results are:

\medskip

({\romannumeral1}) The radio-quiet objects show both long- and
short-term UV/optical continuum variability. No evidence has been found
for periodic behaviour. The continuum typically varies by less than a
factor of two on long (weeks -- months) timescales, with the variability
amplitude rapidly decreasing towards shorter timescales. This trend can
be quantified in terms of a fluctuation power density spectrum (PDS), in
which the power P$(f) \propto f^{\alpha}$ where $f$ is the temporal
frequency. The observed PDS spectral index $\alpha$ is about $-2$ in the
UV for NGC5548 (Krolik {\it et al.}, 1991) and $-2.5$ for NGC4151
(Edelson {\it et al.}, 1996). There is no indication of any
high-frequency cutoff in the UV/optical PDS -- significant variability is
seen down to the temporal sampling limit, which for NGC4151 is only a few
hours. Such high sampling rates correspond to the light-crossing
time of a few tens of Schwarzschild radii around a $\sim 10^7$
M$_{\odot}$ black-hole. At low temporal frequencies (corresponding to a
year or so) the PDS must turn-over to be consistent with the long-term
variability trends.

\smallskip

({\romannumeral2}) The variability amplitude decreases towards longer
wavelengths, indicating a hardening of the UV/optical continuum when
brighter. This relation is difficult to quantify due to contamination by
starlight and weak broad emission lines at optical wavelengths, although
a correction was applied for this when thought significant. For
NGC5548, NGC3783 and Fairall~9 the continuum at 1350\AA\ varies by a
factor of about 3 more than at 5100\AA\ (Reichert {\it et al.}, 1994;
Korista {\it et al.}, 1995; Santos-Lleo {\it et al.}, 1996). For NGC4151
the factor is higher (6--7). No such trend was seen in the intensive
multi-waveband study of the BL~Lac object PKS~2155$-$304 (Edelson {\it et
al.} 1995).

\smallskip

({\romannumeral3}) No clear time-delay or lag has been found between the
continuum variations in the UV/optical. For NGC5548, NGC3783, NGC4151 and
Fairall~9 the lags derived using a variety of cross-correlation
techniques are consistent with zero. An example result for NGC5548 using
HST (1350\AA) and ground-based optical (5100\AA) data is shown in
Figure~1, computed by the ZDCF method (Alexander 1997). The optical
continuum lags the UV by $< 0.8$ days, consistent with the 1.2 day limit
derived by Korista {\it et al.} (1995) using other methods. Similarly,
Edelson {\it et al.} (1996) derived a lag of $<1$ day between 1275 and
5125\AA\ for NGC4151, and Reichert {\it et al.} (1994) find a lag of $<2$
days between 1460 and 5000\AA\ for NGC3783. The upper limits on the lags
between the various UV continuum bands are somewhat smaller ({\it e.g.}
$<0.15$ days for NGC4151).

\medskip

The latter result is perhaps the most important and has been the subject
of intensive observational and theoretical study during the last few
years. The individual limits depend in part on the actual sampling rate
of the light-curves, which in hindsight were too low for some of the
early campaigns. Increasing the sampling rate has been one of the prime
drivers behind the more recent campaigns. The NGC5548 HST campaign
achieved 1 day sampling for 39 days. The NGC4151 IUE campaign achieved
$\approx 70$ minute sampling for 9.3 days. The recent NGC7469 IUE
campaign achieved an average rate of $\sim 5$ hours for $\sim50$ days.
The observing time requirements of such campaigns severely limit the
number of AGN which can be monitored without access to dedicated
monitoring facilities.

\subsection{UV, EUV, X-ray and $\gamma$-ray Continuum Variability}

The IUE and HST campaigns provide information on the UV continuum
($>1150$\AA), but the need to probe shorter wavelengths is clear. The
high-energy continuum is where a substantial, perhaps dominant, fraction
of the bolometric luminosity emerges and is believed to be the continuum
photoionizing the emission-line gas. Monitoring the high-energy continuum
is more difficult due to the limited availability of observing facilities
and their usually lower sensitivity compared to the UV/optical. However,
the AGN Watch has successfully conducted several recent campaigns to
investigate in particular the relation between the UV/optical and
EUV/X-ray/$\gamma$-ray continua.

\subsubsection{NGC5548}Following the discovery of simultaneous
UV/optical continuum variability in NGC5548 (Clavel {\it et al.}, 1991),
another consortium monitored the object with IUE and Ginga (Clavel {\it
et al.}, 1992). They found a strong correlation between the UV and hard
(2--10 keV) X-ray continua during 1989--90, the variations 
occurring simultaneously with a formal limit on any lag of $\pm6$ days. They
noted, however, that there are other epochs when the UV appears stronger
than expected from the correlation observed in 1989--90. This behaviour
strongly suggests that more than one continuum emission mechanism
contributes significantly in the UV and optical wavebands in NGC5548.

During the intensive HST, IUE and ground-based campaign on NGC5548 in
1993, AGN Watch also performed an EUVE monitoring campaign (Marshall {\it
et al.}, 1997). The deep survey imaging detector light-curve indicates
large, rapid variability in the 50--100\AA\ band. The EUV continuum varied
by a factor of four over a few days and by a factor of two in
less than a day. In the most intensively sampled period which overlaps
with the HST campaign, the EUV continuum has a variability amplitude
twice that of the UV at 1350\AA, but they vary simultaneously to within
$<0.25$ days. The short overlap in time (8 days) of these campaigns
preclude a search for long lags, but taken at face value these results
extend the UV and optical continuum correlations noted above into the EUV
-- the variability amplitude increases towards shorter wavelengths and
there is no detectable wavelength-dependent lag. Marshall {\it et al.}
note that similar, but even larger, EUV relative to UV variations were
also seen in Mrk478 (Marshall {\it et al.}, 1995).

\subsubsection{NGC4151}The AGN watch performed it's
most comprehensive monitoring campaign up to that date in December 1993
on the low luminosity Seyfert~1 galaxy, NGC4151. This object has been the
subject of many studies, but the new campaign uniquely obtained
quasi-simultaneous data covering more than 5 decades in energy during
a 10 day period. This campaign is described in detail in these
proceedings by Edelson, so only the conclusions will be given here.

The UV/optical continuum variability characteristics are described above.
The high-energy continuum variability measured using ROSAT, ASCA and CGRO
is complex (Warwick {\it et al.}, 1996; Edelson {\it et al.}, 1996). The
$<1$ keV continuum did not vary significantly. The 1--2 keV continuum
varied simultaneously with the UV continuum at 1275\AA\ to within
0.3~days, but by a factor of 3 more in amplitude. Early in the campaign
the 1--2 keV continuum increased by a factor of 1.45 in just 2 days. The
50--150 keV continuum varied with a similar amplitude to the UV, but
there is no clear relation between the form of the $\gamma$-ray and
low-energy light-curves.

A clear correlation was found between the 2-10 keV absorption corrected
X-ray flux and the UV in 1993 (Warwick {\it et al.}, 1996), with a
similar slope to that found in an earlier study using EXOSAT and IUE data
by Perola {\it et al.} (1986). However, the normalization of the
correlation is different, with a relatively stronger UV continuum
in 1993 implying, as found for NGC5548, that another emission mechanism
contributes significantly in the UV and optical wavebands.

\begin{figure}[ht]
\psfig{figure=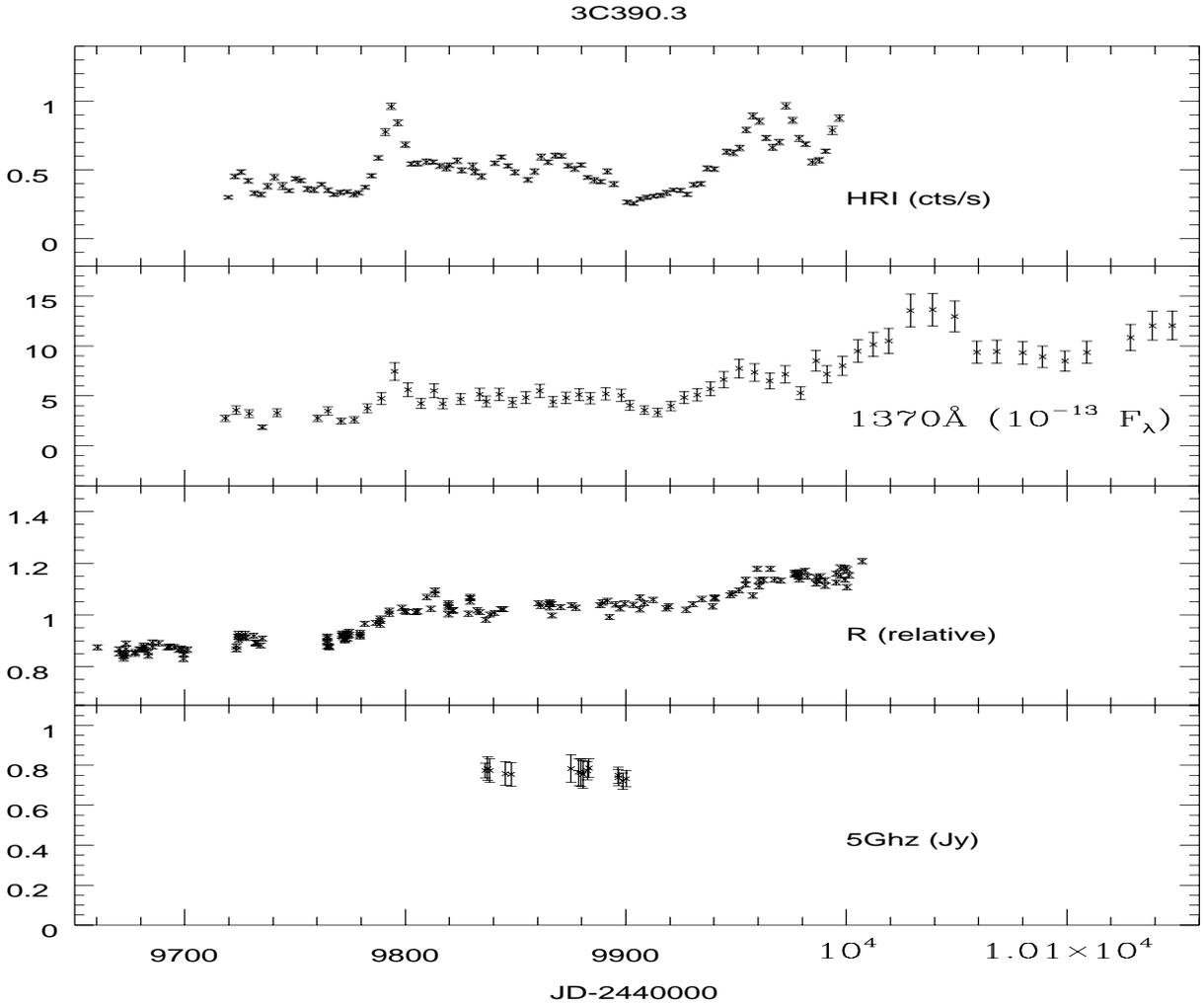,width=18cm,height=14cm,angle=0}
\caption{Continuum data for 3C390.3 in the X-ray (ROSAT/HRI), UV (IUE),
R-band and Radio (MERLIN) obtained during the AGN Watch campaign in 1995--96.}
\end{figure}

\section{THE BROAD LINE RADIO GALAXY 3C390.3}

To complement it's earlier campaigns on radio-quiet Seyfert~1 galaxies,
the AGN Watch monitored the luminous Broad Line Radio Galaxy (BLRG)
3C390.3 during 1995--96. This classical FR~II radio-galaxy is one of the
nearest ($z=0.056$) superluminal sources (Alef {\it et al.}, 1988). It
displays a complex broad emission-line spectrum, particularly
double-peaked Balmer lines usually taken as the signature of an accretion
disk or bi-conical structure ({\it e.g.} Zheng {\it et al.}, 1991). These
properties combined with a history of large amplitude continuum
variations made it a prime candidate for intensive multi-waveband
monitoring. 

We obtained $\sim 1.5$ ksec ROSAT HRI (0.1--2 keV) observations every 3
days for 9 months, IUE observations averaging every 6 days for 14 months,
and numerous optical observations. A 3 month long radio campaign at 5~GHz
using the MERLIN array was also undertaken halfway through the X-ray
campaign. Unfortunately, the IUE campaign was terminated prematurely in
March 1996 due to technical problems with the satellite. We note that
3C390.3 is a factor of 10--50 fainter in observed UV flux than the
previously monitored Seyfert~1 galaxies, so even with longer integration
times ($\sim 6$ hours), the IUE spectra are of significantly lower
quality than in the previous campaigns. In addition only IUE/SWP spectra
(1150--1950\AA) were obtained so there is little constraint on the UV
spectral shape. These multi-waveband data are currently being analysed,
so only initial results will be presented here. The X-ray (HRI counts per
second), UV (mean 1340--1400\AA\ flux in 10$^{-13}$ erg cm$^{-2}$
s$^{-1}$ \AA$^{-1}$), R-band (relative flux converted from magnitudes)
and 5Ghz (core flux in Jy) light-curves are shown in Figure~2, using data
from Leighly {\it et al.} (1997), O'Brien {\it et al.} (in prep.) and
Dietrich {\it et al.} (in prep.).

\begin{figure}[ht]
\psfig{figure=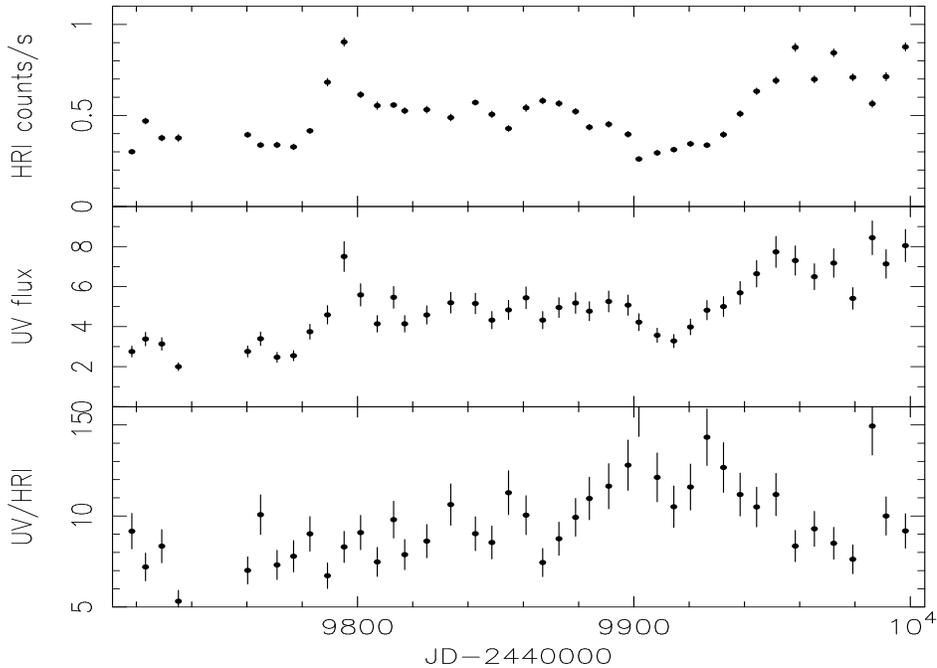,width=18cm,height=10cm,angle=270}
\caption{The UV and re-sampled HRI data for 
3C390.3. HRI data in units of counts/s and UV in
10$^{-13}$ erg cm$^{-2}$ s$^{-1}$ \AA$^{-1}$. The UV/HRI ratio is shown in
the lower box.}
\end{figure}

\subsection{X-ray and UV Continuum Variability}

The X-ray and UV light-curves indicate strong variability, with
normalized variability amplitudes (NVA; Edelson {\it et al.} 1996) in the
period of overlap of 0.32 and 0.31 respectively. The X-ray and UV light
curves show a similar functional form and the cross-correlation function
obtained using the ZDCF method gives a lag of less than 3 days, and is
consistent with zero. The UV continuum at 1370\AA\ and 1855\AA\ also vary
in phase. However, there are some differences in detail between the X-ray
and UV continua. The X-ray flare at JD~2449795 and the low-state X-ray
`dip' beginning around JD~2449900 are less pronounced in the UV. This is
shown more clearly in Figure~3, where the UV and X-ray (re-sampled to the
UV temporal spacing) light-curves are plotted together with their ratio,
which is higher during the X-ray dip.

\begin{figure}[ht]
\psfig{figure=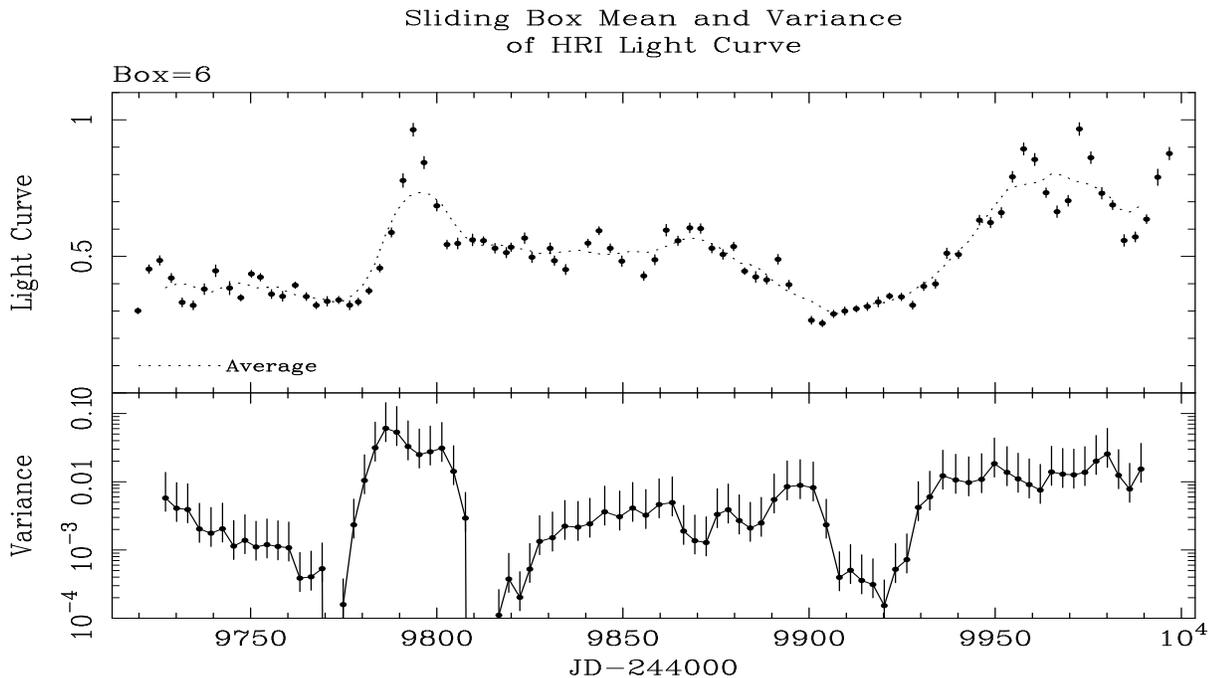,width=18cm,height=10cm,angle=270}
\caption{The HRI data for 3C390.3. The top curve shows the original data
(counts/s). A sliding box-mean (box width 6) is superimposed as the dashed
line whilst the lower curve shows the variance ((counts/s)$^2$) calculated over
the same sliding box.}
\end{figure}


The similar X-ray and UV NVA is in contrast to that observed for NGC4151,
but is consistent with the NVA pattern observed in the BL Lac object
PKS~2155$-$304. However, powerlaw spectral fitting to two ASCA spectra
obtained during the HRI campaign on JD~2449732.5 and 2449842.5 show
photon index variability (Leighly {\it et al.}, 1997). The X-ray spectrum
became softer ($\Delta\Gamma \sim 0.1$) when the 2--10 keV flux increased
by a factor of 1.5, a trend consistent with that generally seen from
Seyfert 1 galaxies. The significant X-ray spectral variability implies
that although the NVA is similar in the soft X-ray and UV wave-bands,
they cannot arise from the same continuum component. There is no evidence
for a soft X-ray excess component in the ASCA spectra, thus the ROSAT HRI
data are sampling the X-ray powerlaw continuum component.

The 3C390.3 X-ray light-curve is uniquely well sampled on time-scales
from days to months, and looks different to those obtained for AGN on
shorter time-scales ({\it e.g.} McHardy and Czerny, 1987). The
combination of flares interspersed among quiescent periods (see Figure~4)
is evidence for non-linear behaviour (Leighly and O'Brien, 1997). Thus, the
3C390.3 X-ray light-curve probably cannot be modeled by a superposition of
independent events, such as in shot-noise models associated with flares in
the disk--corona, or spots on the disk modulated by rotation and
gravitational effects ({\it e.g.} Abramowicz et al. 1991).

\subsection{Optical Continuum Variability}

The NVA is 0.10 for the R-band, and is slightly higher at V and lower at
I (Dietrich {\it et al.}, in prep.). Thus, the trend of decreasing NVA with
increasing wavelength seen in in the UV and optical for Seyfert~1
galaxies is also seen in 3C390.3. Formally the cross-correlation
methods give a small lag of 3--9 days for the R-band relative to the UV
or X-ray, but more careful analysis is required to investigate the
significance of this result. The impression from Figure~2 is that the
R-band light-curve is not simply a smeared-out version of the X-ray/UV.
It does not show the early X-ray/UV flare or the long dip, yet follows
some other trends closely, such as the increases around JD~2449800 and
2449940, albeit with lower amplitude.

\newpage

\subsection{Radio Variability}

The 5~GHz radio flux showed no significant variability ($< 2$\%) during a
period when the X-ray continuum varied by $\pm 20$\% (NVA $=0.13$). The
simplest explanation is that the radio flux originates from a region much
larger than the X-ray source. Thus, the X-ray variations need not be
connected with the superluminal components seen at 5~GHz, which appear
some 5 milli-arcseconds ($\sim 10$pc) from the radio core (Alef {\it et al.},
1988). One could argue that the X-ray/UV flare seen early in the campaign
may be associated with a separate continuum component, possibly a
relativistic jet. However, the reduction in X-ray variability before and
after the flare (Figure~4) argues against this, as superimposing another
variable component would be expected to increase the variability
amplitude.

\section{THEORETICAL IMPLICATIONS}

The observational results summarized above, and in particularly those
related to multi-waveband phase-delay and variability amplitude
comparisons,
comprise
the best available data for these non-blazar AGN. Detailed
comparisons between the observations and theoretical models are
presented elsewhere, including several papers in these proceedings, 
but we note a few major points:

\subsection{Accretion Disk Models}

The lack of a detectable UV/optical continuum lag appears inconsistent
with the predictions of standard accretion disk models in which the
signal connecting parts of the disk emitting at different temperatures
travels through the disk ({\it e.g.} Krolik {\it et al.} 1991). The
derived upper-limits imply signal propagation speeds of $>0.1 c$,
significantly faster than the local sound speed or photon diffusion rate.
Edelson {\it et al.} (1996) propose a disk model for NGC4151 in
which all the variable UV/optical emission comes from a small ($<0.1$
light-day), hot inner region. This model predicts UV/optical lags and NVAs
in agreement with the observations, but does not directly explain
the simultaneous large-amplitude soft X-ray variations.

\subsection {Reprocessing Models}

The lack of a phase delay between the X-ray, UV and optical continua
imply that the entire optical -- X-ray continuum emission, and possibly
the $\gamma$-ray, are closely inter-related. The hypothesis is that one
band provides the `seed photons' which are then `reprocessed' by some
mechanism to drive the emission in another. Which band provides the seed
photons and which the reprocessed is a difficult problem, and feedback
between the bands is also likely. However, the different proposed
reprocessing mechanisms, such as thermal reprocessing of X-ray photons to
UV/optical by an accretion disk or Thomson thick clouds, or Compton
scattering of UV photons to X-ray by hot electrons around a disk predict
different behaviour ({\it e.g.} Nandra, 1994). Simultaneous variability
data should provide the cleanest test between the proposed mechanisms.

\subsubsection{Seyfert~1 Galaxies}The X-ray spectral features
seen in Seyfert~1 galaxies are consistent with those predicted by models
in which X-ray photons are reprocessed by dense gas into
UV and optical photons (Mushotzky, Done and Pounds 1993 and references
therein). These features include an Fe K$\alpha$ emission-line due to
fluorescence and a hard (few tens of keV) X-ray `tail' due to Compton
reflection. Most of the incident X-ray energy, however, is absorbed by the
gas and re-radiated thermally at UV/optical wavelengths determined by the
effective temperature of the gas ($\sim 10^{5}$ K). The X-rays would therefore
synchronise variability between the wave-bands, in agreement with the lack
of observed phase-delay. In practice, several variations on the theme of
disk-corona models or Thomson thick cloud models have been proposed,
which by adjusting factors such as the patchiness of the disk-corona, the
cloud covering factor and the Thomson optical depth of the gas can
produce a variety of spectra. Even for the best studied source, NGC4151,
a consensus has not yet been reached ({\it e.g.} Warwick {\it et al.}
1996; Poutanen {\it et al.} 1996; Zdziarski and Magdziarz 1996). The
excess UV emission seen in NGC5548 and NGC4151 at some epochs must also
be explained, for example by invoking significant intrinsic accretion
disk emission due to viscous heating.

\subsubsection{3C390.3}
The X-ray/UV/optical continuum variability characteristics seen in
3C390.3 appear quite similar to those seen in the radio-quiet AGN, so the
X-ray reprocessing model seems an attractive proposition for this object.
A 1993 ASCA X-ray spectrum of 3C390.3 (Eracleous {\it et al.}, 1996)
shows evidence for a broad Fe K$\alpha$ line, believed to be evidence for
X-ray reprocessing by gas near a black-hole (Tanaka {\it et al.}, 1995).
Their ASCA GIS3 spectrum also show evidence for a hard X-ray tail,
consistent with Compton reflection; however, calibration uncertainties
could affect this result. Determining whether the difference in the
relative NVA in the X-ray and UV between 3C390.3 and NGC4151 is related
to NGC~4151 being in an unusually high UV state during the 1993 campaign,
or reflects a more intrinsic difference requires more detailed study. The
differences between the X-ray/UV/optical light-curves in 3C390.3 must
also be explained.


\section{FUTURE REQUIREMENTS}

Despite much effort the kind of intensive multi-waveband campaigns
discussed in this paper are few and far between. Much organizational
effort is involved in conducting such campaigns over many weeks,
requiring co-operation from schedulers and other observers as well as
taking up valuable research time for many astronomers. The size of
AGN watch is testament to these facts, and it is not a practical way to
proceed over the long-term. In addition, very few AGN are monitored
leading to poor statistics, made even worse by the knowledge that to gain
observing time we must pre-select AGN known to be highly variable, which
may introduce significant and unknown biases into the results.
For the future there appear to be two approaches: (1) Multi-waveband
facilities which allow simultaneous observations across a wide frequency
range without requiring huge collaborations; (2) Robotic facilities,
either ground- or space-based, which provide probably limited frequency
access, but require little effort for scheduling. 

An example of the first approach is the forthcoming ESA X-ray
Multi-Mirror Observatory (XMM). XMM will not only carry a large X-ray
(0.1--10 keV) telescope system but also a small (30cm), but sensitive,
co-aligned UV/optical telescope -- the Optical Monitor (XMM-OM). The
XMM-OM will provide both broad-band filter and grism capability from
1500--6000\AA. Thus, XMM will allow rapid, simultaneous X-ray and
UV/optical monitoring of every AGN pointed at in guest-observer mode, and
others in the surrounding field of view. For example, for a bright AGN
such as 3C390.3 a 10 minute observation will give an $\approx 100\sigma$
detection in the X-ray and $\approx 50\sigma$ in the 1700--2300\AA\
filter. X-ray and UV/optical grism spectra can also be obtained on longer
timescales.

The second approach is already under-way for ground-based robotic
telescopes. Exiting facilities are fairly small, but plans for 1-2m class
facilities are well advanced (Parker 1996). In space one possible
candidate is Lobster-eye, an all-sky, soft X-ray (0.5--2.4 keV)
monitoring satellite (Priedhorsky {\it et al.}, 1996). This SMEX-class
mission proposal uses advanced micro-channel plate technology to image a
large sky area ($\sim 2\pi$ sr) simultaneously, spinning the satellite to
get full coverage. The base-line study gives a daily $5\sigma$ detection
limit of $2\times10^{-12}$ erg cm$^{-2}$ s$^{-1}$. This would result in
daily observations of $\sim1000$ AGN, with better sampling and/or quality
for the brighter objects. Such X-ray light-curves, combined with data from
other wave-bands, would totally revolutionise AGN continuum studies.

\acknow
We are grateful to Mathias Dietrich for the optical data of 3C390.3,
Karen Wills for the MERLIN data and Tal Alexander for the ZDCF code. The
AGN Watch also gratefully acknowledges the numerous observatory
directors, schedulers and TACs whose facilities we have used.


\begin{references}

Abramowicz, M.A., Bao, G., Lanza, A. and Zhang, X.-H. {\it A\&A}, 245, 454 
(1991).

Alef, W. G\"otz, M.M.A., Reuss, E. and Kellermann, K.I., {\it A\&A}, 192, 53
(1988). 

Alexander, T., in {\it Astronomical Time Series}, eds. Leibowitz, E.,
Maoz, D., Sternberg, A., Kluwer, Dordrecht, in press (1997)

Clavel, J.C., Reichert, G.A., Alloin, D., Crenshaw, D.M., Kriss, G. 
{\it et al.}, {\it ApJ}, 366, 64 (1991).

Clavel, J.C., Nandra, K., Makino, F., Pounds, K.A. {\it et al.}, {\it
ApJ}, 393, 113 (1992).

Crenshaw, D.M., Rodr\'{\i}guez-Pascual, P.M., Penton, S.V., Edelson, R.A., 
Alloin, D. {\it et al.}, {\it ApJ}, 470, 322 (1996).

Dietrich, M., Kollatschny, W., Peterson, B.M., Bechtold, J., Bertram, R. 
{\it et al.}, {\it ApJ}, 408, 416 (1993).

Edelson, R.A., Krolik, J., Madejski, G., Maraschi, L., Pike, G. {\it et
al.}, {\it ApJ}, 438, 120 (1995).

Edelson, R.A., Alexander, T., Crenshaw, D.M., Kaspi, S., Malkan, M.A.
{\it et al.}, {\it ApJ}, 470, 364 (1996).

Eracleous, M., Halpern, J.P. and Livio, M., {\it ApJ}, 459, 89 (1996).

Kaspi, S., Maoz, D., Netzer, H., Peterson, B.M., Alexander, T. {\it et
al.}, {\it ApJ}, 470, 336 (1996).

Korista, K.T., Alloin, D., Barr, P., Clavel, J., Cohen, R.D. 
{\it et al.}, {\it ApJS}, 97, 285 (1995).

Krolik, J.H., Horne, K., Kallman, T.R., Malkan, M.A., Edelson, R.A., and
Kriss, G.A., {\it ApJ}, 371, 541 (1991).

Leighly, K.M. and O'Brien, P.T., {\it ApJL}, in press (1997)

Leighly, K.M., O'Brien, P.T., Edelson, R., George, I.M., Malkan, M.A.
{\it et al.}, {\it ApJ}, in press (1997)

Marshall, H.L., Carone, T.E., Shull, J.M., Malkan, M.A., and Elvis, M.,
{\it ApJ}, 479, 169 (1995).
 
Marshall, H.L. Carone, T.E., Peterson, B.M., Clavel, J., Crenshaw, D.M.
{\it et al.}, {\it ApJ}, in press (1997).

McHardy, I. and Czerny, B., {\it Nature}, 325, 696 (1987).

Mushotzky, R.F., Done, C., and Pounds, K.A., in {\it Annual Review of
Astronomy \& Astrophysics}, 31, 717-761 (1993).

Nandra, K., in {\it Reverberation Mapping of the BLR in AGN}, eds.
Gondhalekar, B.M., Horne, K., Peterson, B.M., ASP Conference Series, 69,
pp. 273-291 (1994)

Parker, N., in {\it Spectrum}, PPARC Observatories Newsletter No. 10, (1996).

Perola, G.C., Piro, L., Altmore, A., Fiore, F., Boksenberg, A., {\it
ApJ}, 306, 508 (1986)

Peterson, B.M., Balonek, T.J., Barker, E.S., Bechtold, J., Bertram, R. 
{\it et al.}, {\it ApJ}, 368, 119 (1991).

Peterson, B.M.,  Alloin, D., Axon, D., Balonek, T.J., Bertram, R. 
{\it et al.}, {\it ApJ}, 392, 470 (1992).

Peterson, B.M., Berlind, P., Bertram, R., Bochkarev, N.G., Bond, D. 
{\it et al.} {\it ApJ}, 425, 622 (1994).

Poutanen, J., Sikora, M., Begelman, M.C. and Magdziarz, P., {\it ApJ},
465, 107L (1996).

Priedhorsky, W.C., Peele, A.G. and Nugent, K.A., {\it MNRAS}, 279, 733
(1996).

Reichert, G.A., Rodr\'{\i}guez-Pascual, P.M., Alloin, D., Clavel, J.,
Crenshaw, D.M., Kriss, G.A. {\it et al.}, {\it ApJ}, 582, 608 
(1994).

Rodr\'{\i}guez-Pascual, P.M., Alloin, D., Clavel, J., Crenshaw, D.M., 
Horne, K. {\it et al.}, {\it ApJ}, submitted (1997).

Stirpe, G.M., {\it et al.}, {\it ApJ}, 425, 609 (1994).

Tanaka, Y, Nandra, K., Fabian, A.C., Inoue, H. and Otani, C., {\it
Nature}, 375, 659 (1995).

Warwick, R.S., Smith, D.A., Yaqoob, T., Edelson, R., Johnson, W.N. 
{\it et al.}, {\it ApJ}, 470, 349 (1996).

Zdziarski, A.A. and Magdziarz, P., {\it MNRAS}, 279, L21 (1996).

Zheng, W., Veilleux, S. and Grandi, S.A., {\it ApJ}, 381, 411 (1991).

\end{references}
\end{document}